\documentclass[prd, twocolumn, nofootinbib, floatfix]{revtex4-1}
\usepackage{amsmath}
\usepackage{graphicx}
\usepackage{dcolumn}
\usepackage{bm}
\usepackage{epsfig}
\usepackage{amssymb,latexsym,mathrsfs}
\usepackage{graphicx}
\usepackage{color}
\usepackage{hyperref}

\def \beq  {\begin{equation}}
\def \eeq  {\end{equation}}
\def \ber  {\begin{eqnarray}}
\def \eer  {\end{eqnarray}}

\hypersetup{
    colorlinks=true,
    linkcolor=red,
    citecolor=blue,
}

\newcommand{\be}{\begin{equation}}
\newcommand{\ee}{\end{equation}}
\newcommand{\bea}{\begin{eqnarray}}
\newcommand{\eea}{\end{eqnarray}}

\newcommand{\om}{\Omega_m}
\newcommand{\omt}{\Omega_{m,0}}

\newcommand{\sigf}{\sigma_f}
\newcommand{\sige}{\sigma_8}
\newcommand{\sigt}{\sigma_{8,0}}
\newcommand{\lcdm}{$\Lambda$CDM}

\begin{document}
\title{Model Independent Tests of Cosmic Growth vs Expansion} 
\author{Arman Shafieloo$^{1,2,3}$, Alex G.\ Kim$^4$, Eric V.\ Linder$^{2,4,5}$} 
\affiliation{$^1$  Asia Pacific Center for Theoretical Physics, Pohang, Gyeongbuk 790-784, Korea}
\affiliation{$^2$ Institute for the Early Universe WCU, Ewha Womans University,
Seoul, 120-720, Korea }
\affiliation{$^3$ Department of Physics, POSTECH, Pohang,Gyeongbuk 790-784, Korea}
\affiliation{$^4$ Lawrence Berkeley National Laboratory, Berkeley, CA 94720, 
USA} 
\affiliation{$^5$ Berkeley Center for Cosmological Physics, 
University of California, Berkeley, CA 94720, USA}

\begin{abstract}
We use Gaussian Processes to map the expansion history of the universe 
in a model independent manner from the Union2.1 supernovae data and then 
apply these reconstructed results to solve for the growth history.  By 
comparing this to BOSS and WiggleZ large scale structure data we examine 
whether growth is determined wholly by expansion, with the measured 
gravitational growth index testing gravity without 
assuming a model for dark energy.  A further model independent analysis 
looks for redshift dependent deviations of growth from the general 
relativity solution without assuming the growth index form.  Both approaches 
give results consistent with general relativity.
\end{abstract}

\date{\today} 

\maketitle

\section{Introduction} 

Cosmological surveys have advanced to provide precision determinations 
of the distance-redshift relation and, to a lesser extent, the 
growth-redshift relation for cosmic history at redshifts $z\lesssim1$.  
These improvements allow not only determination of cosmological parameters 
but fundamental tests of the cosmological framework in a more model 
independent manner.  Rather than assuming a model with cold dark matter plus 
a cosmological constant (\lcdm) or plus dark energy parametrized by a 
constant equation of state ratio $w$ or time varying $w(z)=w_0+w_a z/(1+z)$, 
one might like to investigate the expansion history $H(z)$ and growth 
factor $D(z)$ or the growth rate $f(z)=d\ln D/d\ln (1+z)$ directly, with 
minimal assumptions. 

Reconstruction of the expansion history, in terms of the inverse Hubble 
parameter $H^{-1}(z)$ or deceleration parameter 
$q(z)=-d\ln H^{-1}/d\ln (1+z)-1$, can be carried out purely kinematically, 
without assuming a particular theory of gravity or field equations (i.e.\ 
Friedmann equations).  Gaussian Processes \cite{gpml} prove to be an 
effective statistical technique for carrying out such a reconstruction 
from distance data, as done in \cite{skl}. 

Growth of matter density perturbations into large scale structure, 
however, depends explicitly on the dynamics, i.e.\ the gravitational 
force law.  Within general relativity (and pressureless matter being the 
only significantly clustering component), expansion and growth are 
locked together, either one determining the other.  Given that recently 
growth data have advanced to cover a reasonable redshift range, 
$z\approx0-0.8$, at $\sim10\%$ precision, it is interesting to test whether 
this interrelation actually holds.  We can enlarge the Gaussian Process 
technique to do this in a model independent manner (rather than assuming 
a dark energy parametrization), although somewhat less generally than the 
previous expansion history reconstruction in that we must separate out the 
matter density. 

In Sec.~\ref{sec:methods} we briefly review the cosmographic reconstruction 
of expansion history from distance data and describe the extraction of 
growth history from large scale structure data, in particular using 
redshift space distortion measurements.  Section~\ref{sec:datasim} carries 
out likelihood analysis of current data and derives confidence 
contours on the matter density and gravitational growth index.  We outline 
future applications and conclude in Sec.~\ref{sec:concl}.

\section{Gaussian Process Method} \label{sec:methods}

\subsection{From Distances to Expansion} 

Distance measurements play an essential role in our understanding of 
the history and contents of the universe.  They have a linear relation to 
(through integration of) the inverse Hubble parameter or expansion rate 
for a spatially flat Roberston-Walker universe as we assume. 
This linearity is an important property for a Gaussian process (GP), since 
the derivative (or integral) of a GP is another GP, making error propagation 
particularly straightforward.  Thus, for a luminosity distance 
\be 
d_l(z)=(1+z)\,\eta(z)=(1+z)\,\int_0^z dz'\,H^{-1}(z') \ , 
\ee 
if we model $d_l$ as a GP then the conformal distance $\eta$ or inverse 
Hubble parameter $H^{-1}$ is one as well. 

Gaussian processes provide a robust statistical method for using 
stochastic data measured 
at certain points (redshifts) and reconstructing the full function 
(distance-redshift relation or inverse Hubble parameter) describing the 
underlying relation, complete with covariances and without assuming a 
specific model for the relation.  
See \cite{gpml} for detailed explanation of their general application, 
and \cite{la1,la2,la3,skl} for specific application to dark energy and 
cosmology (also see \cite{seikel1,seikel2}, though they fix several aspects 
of the GP and distance model, and \cite{nesseris} for a genetic algorithm 
approach).  
Here we follow most closely \cite{skl}. 

Given data $\mathbf{y}$ at a set of points $Z$ we reconstruct the 
underlying function $\mathbf{f}$, or its derivatives, at any set of points 
$Z_1$.  The probability distribution functions are Gaussians described by 
a mean function $\mathbf{m(Z)}$ and covariance matrix $k(Z_i,Z_j)$: 
\begin{widetext}
\begin{equation}
\left[
\begin{array}{c}
\mathbf{y} \\
\mathbf{f} \\
\mathbf{f'} 
\end{array}
\right]
\sim
\mathcal{N}
\left(
\left[
\begin{array}{c}
\mathbf{m(Z)} \\
\mathbf{m(Z_1)} \\
\mathbf{m'(Z_1)} 
\end{array}
\right]
,
\left[
\begin{array}{ccc}
\Sigma_{00}(Z,Z) & \Sigma_{00}(Z,Z_1)  &  \Sigma_{01}(Z,Z_1)  \\
\Sigma_{00}(Z_1,Z) & \Sigma_{00}(Z_1,Z_1)  &  \Sigma_{01}(Z_1,Z_1) \\ 
\Sigma_{10}(Z_1,Z) & \Sigma_{10}(Z_1,Z_1) &  \Sigma_{11}(Z_1,Z_1) 
\end{array}
\right]
\right) ,
\end{equation} 
\end{widetext}
where 
\begin{equation}
\Sigma_{\alpha\beta} = \frac{d^{\left(\alpha+\beta\right)}k(Z_i,Z_j)}{dz_i^\alpha dz_j^\beta} \ , 
\end{equation} 
and a prime indicates $d/dz$.  

The inferred mean and covariance of the functions are given by
\begin{equation}
\left[
\begin{array}{c}
\overline{\mathbf{f}}\\
\overline{\mathbf{f'}} 
\end{array}
\right]
=
\left[
\begin{array}{c}
\mathbf{m(Z_1)}\\
\mathbf{m'(Z_1)} 
\end{array}
\right]
+
\left[
\begin{array}{c}
\Sigma_{00}(Z_1,Z) \\
\Sigma_{10}(Z_1,Z) 
\end{array}
\right]
\Sigma_{00}^{-1}(Z,Z)  \,\mathbf{y}
\end{equation}
\begin{widetext}
\begin{equation}
\mbox{Cov}\left(
\left[
\begin{array}{c}
\mathbf{f}\\
\mathbf{f'} 
\end{array}
\right]
\right)
=
\left[
\begin{array}{cc}
\Sigma_{00}(Z_1,Z_1) & \Sigma_{01}(Z_1,Z_1) \\ 
\Sigma_{10}(Z_1,Z_1) & \Sigma_{11}(Z_1,Z_1) 
\end{array}
\right]
-
\left[
\begin{array}{c}
\Sigma_{00}(Z_1,Z) \\
\Sigma_{10}(Z_1,Z) 
\end{array}
\right]
\Sigma_{00}^{-1}(Z,Z)
\left[\Sigma_{00}(Z,Z_1),\Sigma_{01}(Z,Z_1)\right].
\end{equation}
\end{widetext}

For the GP covariance function we use a common form, the squared 
exponential, 
\be 
k(z,z')=\sigf^2\,\exp{\left(-\frac{|z-z'|^2}{2l^2}\right)}, 
\ee 
where $\sigf$ defines the overall amplitude of the correlation 
and $l$ measures the coherence length of the correlation.  The parameters 
$\sigma_f^2$ and $l$ are hyperparameters in the fit. 

If $f(z)$ is the reconstructed distance then $f'$ is the reconstructed 
inverse Hubble parameter.  Within general relativity, $H^{-1}$ also 
determines the linear growth history of large scale structure.  While 
the growth factor or growth rate will not be linear functions of $H^{-1}$, 
and so are not GPs themselves, the error propagation is still direct.  The 
basic approach is that supernova distance data allow (model independent) GP 
reconstruction of $H^{-1}(z)$, with its covariances between redshifts, 
as in \cite{skl}, and then this can be propagated to predictions of 
growth.  These can then be compared to growth data from galaxy redshift 
surveys.

\subsection{From Expansion to Growth} 

The linear growth factor is difficult to measure directly, free from 
astrophysical effects such as galaxy bias.  Weak gravitational lensing 
data, which does not involve galaxy bias, is not currently sufficiently 
accurate to be useful for the desired reconstruction.  Therefore we use 
galaxy redshift survey measurements of the 
growth rate through redshift space distortions, whose anisotropic angular 
dependence allows separation from galaxy bias. 

Redshift space distortions arise as follows.  
The matter density perturbations forming large scale structure induce 
gravitational potential inhomogeneities, and these in turn give rise 
to motions of the matter, or peculiar velocities.  These velocities add 
to the galaxy redshift due to cosmic expansion, causing an anisotropic 
observed density field, in redshift space.  Since the peculiar velocities 
are proportional to the growth rate $f=d\ln D/d\ln a$, where the scale 
factor $a=1/(1+z)$, then these redshift 
space distortions can be used as a cosmological probe \cite{hamilton}.  In 
the linear perturbation limit, \cite{kaiser} showed the observed (redshift 
space) galaxy power spectrum is related to isotropic real space density 
power spectrum by 
\be 
P^s(k,\mu)=(b+f\mu^2)^2\,P^r(k) \ , 
\ee 
where $k$ is the wavemode of the density perturbation, $\mu$ is the 
cosine of its angle with respect to the line of sight, and $b$ is the 
galaxy bias. 

Since the power spectrum is proportional to the square of the mass 
fluctuation amplitude, $P^r(k,a)\propto \sige^2(a)\propto D^2(a)$, then 
the redshift space distortion observable is $f\sige\propto dD/d\ln a$ 
(see, e.g., \cite{percwhite}).  Normalized to the present mass fluctuation 
amplitude $\sigt$, an excellent approximation to the 
cosmological and gravitational dependence of this quantity is 
\be 
\phi(a)\equiv \frac{f\sige}{\sigt}=\om(a)^\gamma \, 
e^{\int_a^1 d\ln a'\,[\om(a')^\gamma-1]} \ , 
\ee 
where $\gamma$ is a constant called the 
gravitational growth index \cite{lin05}.  For general relativity and \lcdm, 
$\gamma=0.55$.  The gravitational growth index form has been shown to be 
accurate at the 0.1\% level for a wide variety of dark energy and gravity 
models \cite{lin05,lincahn}, so long as the gravitational strength remains 
scale independent and no strong clustering of dark energy occurs. 

This form immediately allows us to carry out a test of gravity without 
choosing a dark energy model or parametrizing the Hubble expansion, since 
\be 
\om(a)^\gamma=\left[\Omega_{m,0}\,a^{-3}\,\left(H^{-2}/H_0^{-2}\right)\right]^\gamma \ . 
\ee 
The normalization relative to today, i.e.\ the $\sigt$ in $\phi$, ensures 
that there is no dependence on $H(z)$ for redshifts higher than the highest 
growth measurements (and hence distance data, since these extend further).  
We can thus use the model independent GP reconstruction of 
$H^{-1}(a)/H_0^{-1}$ 
from the SN distance data, propagate it to predictions of the growth 
relation, and by comparing to the growth data fit for the matter 
density today $\omt$ and $\gamma$, the latter testing gravity. 

In a second approach, one can actually enhance the model independence by 
writing 
\be 
\phi(a)=\phi_{\rm GR}(a)+\delta\phi(a) \ , 
\ee 
where $\phi_{\rm GR}$ fixes $\gamma=0.55$ as from general relativity, and 
use Gaussian processes to reconstruct the function $\delta\phi$ without 
assuming a functional form given by the growth index $\gamma$.   
(Mathematically, one uses the GR relation as the mean function in the GP 
and sees if the hyperparameter $\sigma_f$ giving the amplitude of deviations 
is consistent with zero.)  
This allows for exploration of a wider variety of extended gravity 
theories.  We use both approaches in the next section.

\section{Results from Distance and Growth} \label{sec:datasim} 

We apply GP to the Union2.1 compilation of supernova distance data 
\cite{union21} and incorporate the $f\sigma_8(z)$ data from 
the Baryon Oscillation 
Spectroscopic Survey (BOSS \cite{reid}), SDSS DR7 \cite{sdssdr7}, WiggleZ 
\cite{wigglez}, 2dF \cite{2dF}, and 6dF \cite{6dF} galaxy surveys.  
Note one must be 
careful to use the growth data values derived without assuming a 
specific expansion model.  Through a scan over the likelihood 
surface, marginalizing over hyperparameters, we can derive confidence 
contours for $\omt$ and $\gamma$, or study the hyperparameters themselves. 

\begin{figure}[!hbtp] 
\includegraphics[angle=-90,width=\columnwidth]{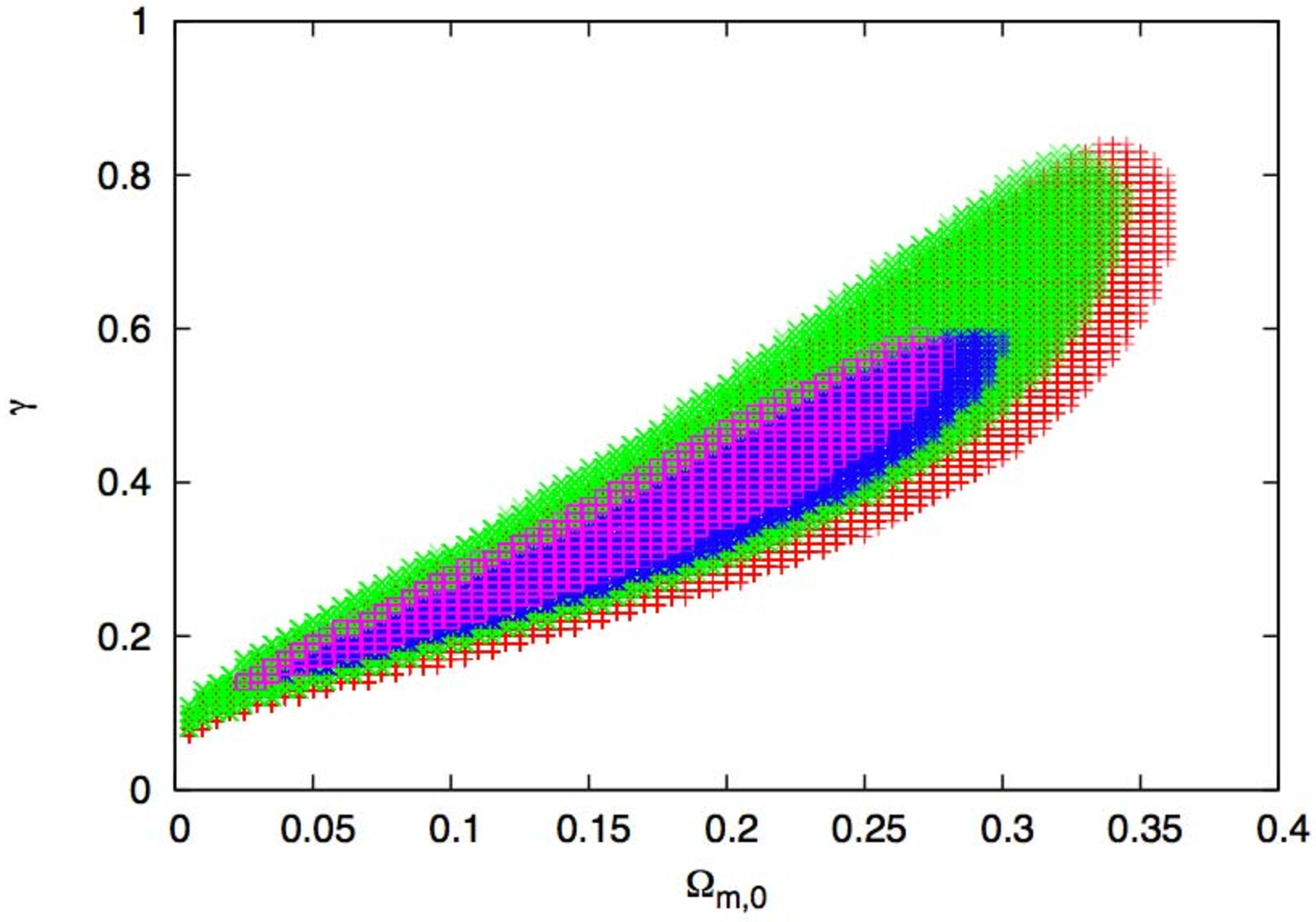} 
\caption{$68\%$ and $95\%$ joint confidence limits on $\gamma$ and 
$\omt$ are shown derived without assuming a dark energy model, using 
current supernovae distance and galaxy clustering growth data.  
The left contour of each pair has $\sigt=0.801$, the WMAP7 concordance 
value \cite{wmap7}, and the right has $\sigt=0.78$ to show the effect 
of a small shift. 
} 
\label{fig:omg}
\end{figure}

Figure~\ref{fig:omg} shows the joint 2D contours on $\omt$--$\gamma$ for 
two different values of $\sigt$.  The results are consistent with GR 
value of $\gamma=0.55$, and the variation of $\sigt$ slides the contours 
in $\omt$ with little effect on the probability distribution of $\gamma$.

The GP method, without assuming any dark energy model, indicates that 
$\gamma$ can take values in a considerable range, though general relativity 
i.e.\ $\gamma=0.55$, is right near the peak of the likelihood.  
In particular, even when fixing $\omt=0.28$, say, values of $\gamma$ as low 
as those characteristic of scalar-tensor theories such as $f(R)$ gravity 
\cite{sotiriou} or as high as that of DGP gravity \cite{dgp}, 0.42 and 0.68 
respectively, are allowed at 95\% CL.  

\begin{figure}[!hbtp] 
\includegraphics[width=\columnwidth]{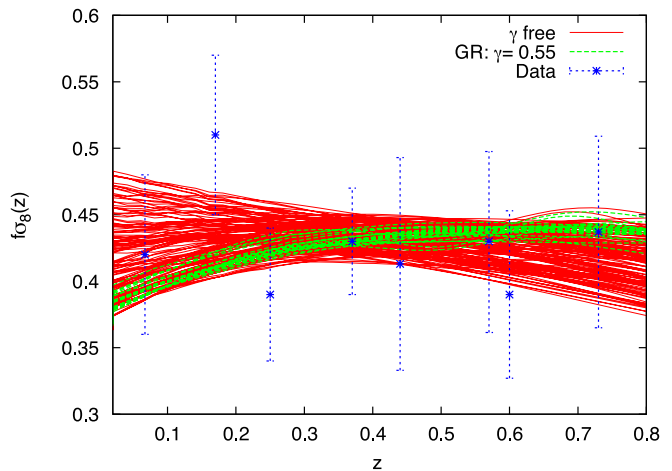} 
\caption{Reconstruction of the growth rate $f\sigma_8(z)$ is shown 
for the case when fixing $\gamma=0.55$ (light green curves) or allowing 
it to float (dark red curves).  Current growth data is overplotted and we 
fixed  $\sigt=0.801$. 
} 
\label{fig:bands}
\end{figure}

Current data therefore does not have the leverage to look for more subtle 
redshift dependent deviations from GR that might not be captured by 
$\gamma$, at least not without assuming a specific model.  We quantify the 
reach of current data in two ways.  Figure~\ref{fig:bands} shows the GP 
reconstructions of the growth rate as a function of redshift, with data 
from BOSS, SDSS DR7, WiggleZ, 2dF, and 6dF overplotted.  
The light green band is composed of samples of reconstructions with 
$\Delta\chi^2<3$ relative to the best fit, when fixing $\gamma=0.55$, 
while the dark red band allows $\gamma$ to float. In both these cases we 
have fixed  $\sigt=0.801$.

\begin{figure}[!t]
\includegraphics[angle=-90,width=\columnwidth]{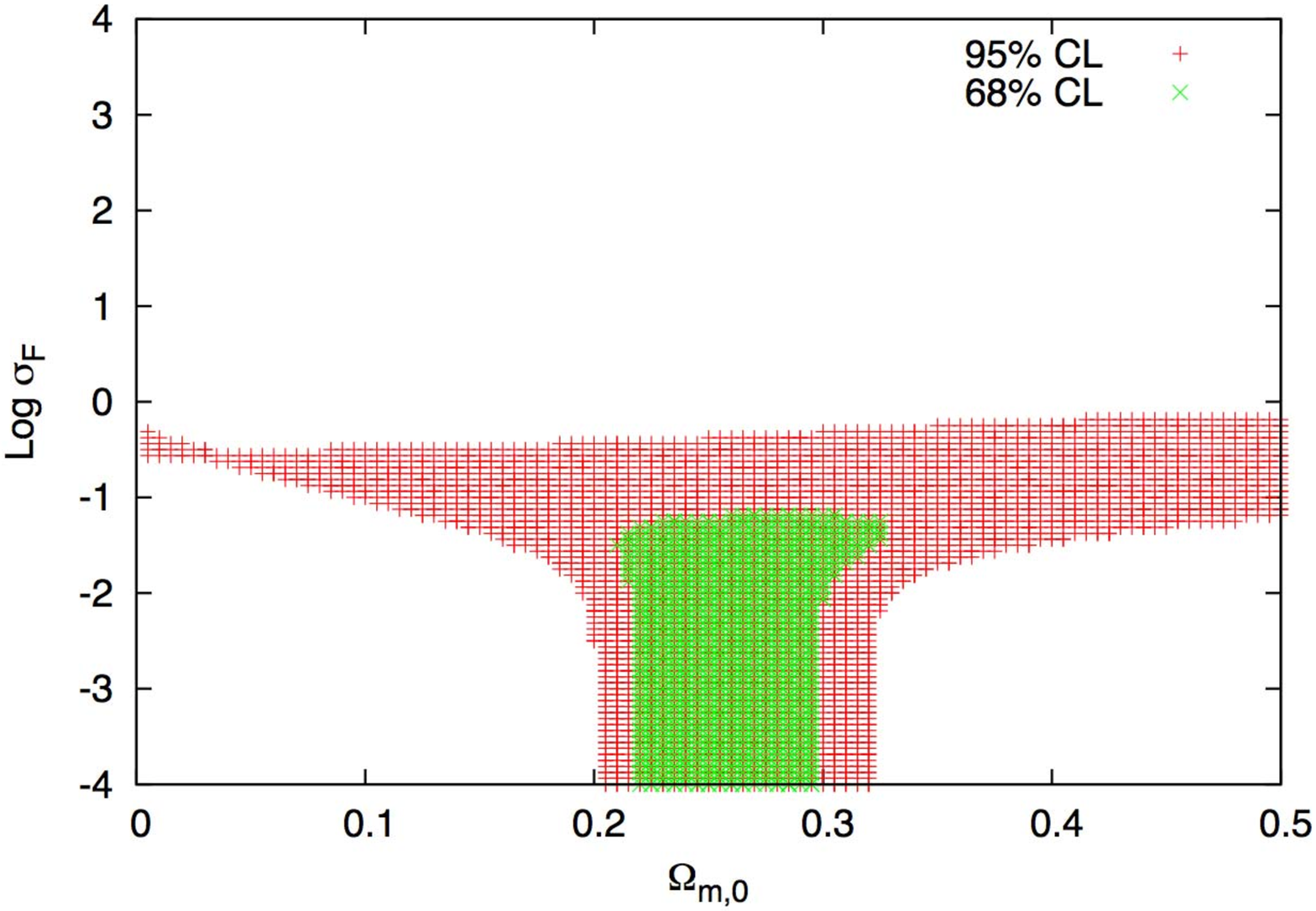}\\ 
\caption{Limits on the amplitude of deviation $\sigf$ from the GR 
growth relation (with model independent dark energy) are shown in 
2D joint confidence contours with the present matter density. 
}
\label{fig:sigf}
\end{figure}

The second method involves taking the even more model independent approach 
of fitting for an arbitrary time dependent correction to the general 
relativity growth rate, 
$\delta\phi(a)=\phi(a)-\phi_{GR}(a)$.  That is, we take GR to provide the 
mean function for the GP and let the data constrain the amplitude of the 
deviations given by the hyperparameter $\sigf$.  Figure~\ref{fig:sigf} 
shows the 2D bound in the $\sigf$--$\omt$ plane. 
The data prefers no deviation from general relativity, i.e.\ $\sigf=0$ is 
within 68\% CL.

\section{Conclusions} \label{sec:concl}

We have demonstrated a method to solve for the expansion and growth histories  
of the universe simultaneously, without assuming any model or parametrization 
for dark energy.  This is a key test of the cosmological framework since 
within Einstein gravity one determines the other.  Using the results we 
derive from Type Ia supernovae and large scale structure data, we test for 
and quantify deviations from general relativity in two ways. 

Gaussian Processes (GP) provide a useful statistical technique for such 
model independent analyses.  The GP reconstruction of the expansion history 
was juxtaposed with growth rate data from redshift space distortion 
measurements in galaxy surveys to obtain probability distributions involving 
the gravitational growth index $\gamma$.  The general relativity value was 
found to be a good fit, although due to uncertainty in the matter density 
$\omt$ and to a lesser extent the mass fluctuation amplitude $\sigt$ a wide 
range of values is tenable within current constraints. 

We further extend the model independence by looking for any deviation in 
the growth rate, without using the growth index formalism.  Building on the 
GP reconstruction we test the growth data for deviations from the prediction 
of general relativity as a function of redshift.  The results are again 
consistent with standard gravity, tested without assuming any particular 
model of dark energy.  Stringent exploration of the laws of gravity, however, 
requires more accurate future growth and distance data.  Ongoing and future 
surveys will greatly enhance our ability to carry out model independent 
investigation of the cosmological framework.

\acknowledgments 

A.S.\ acknowledges 
the Max Planck Society (MPG), the Korea Ministry of Education, Science and 
Technology (MEST), Gyeongsangbuk-Do and Pohang City for the support of the 
Independent Junior Research Groups at the Asia Pacific Center for 
Theoretical Physics (APCTP). This work has been supported by World Class 
University grant R32-2009-000-10130-0 through the National Research 
Foundation, MEST and 
DOE grant DE-SC-0007867 and the Director, Office of Science, Office of 
High Energy Physics, of the U.S.\ Department of Energy under Contract 
No.\ DE-AC02-05CH11231.  
A.G.K.\ thanks 
IEU and APCTP for hospitality during this work.  



\end{document}